\documentclass[twocolumn, amssymb, amsmath, aps, prb, showpacs, 10pt]{revtex4-1}
\usepackage[dvips]{graphicx}
\usepackage[dvips]{color}
\usepackage{hyperref}
\hypersetup{
    colorlinks=true,
    linkcolor=blue,
}

\begin{document}
\title{Mn-site doping and its effect on inverted hysteresis and thermomagnetic irreversibility behavior of antiferromagnetic Mn$_5$Si$_3$ alloy}
\author{S. C. Das}
\author{K. Mandal}
\author{N. Khamaru}
\author{S. Pramanick}
\author{S. Chatterjee}
\email{souvik@alpha.iuc.res.in}
\email{souvikchat@gmail.com}
\affiliation{UGC-DAE Consortium for Scientific Research, Kolkata Centre, Sector III, LB-8, Salt Lake, Kolkata 700 106, India}
\begin{abstract}
The structural and magnetic behavior of Mn-site doped intermetallic manganese silicide alloys of nominal compositions Mn$_{5-x}$A$_x$Si$_3$ ($x$ = 0.05, 0.1, 0.2 and A = Ni, Cr) have been investigated with a focus to the inverted hysteresis behavior and thermomagnetic irreversibility. Room temperature x-ray powder diffraction data confirm that all the doped alloys crystallize in hexagonal $D8_8$ type structure with space group $P6_3/mcm$. The doped alloys are found to show paramagnetic (PM) - collinear antiferromagnetic (AFM2) - noncollinear antiferromagnetic (AFM1) transitions during cooling from room temperature. A significant decrease in the critical values of both AFM1-AFM2 transition temperatures and fields have been observed with the increasing Ni/Cr concentration. Inverted hysteresis loop, field-induced arrest, and thermomagnetic arrest, the key features of the undoped Mn$_5$Si$_3$ alloy, are found to be significantly affected by the Mn-site doping and eventually vanishes with 4\% doping.
\end{abstract}
\maketitle

\section{Introduction}
Apart from the conventional microelectronic application, manganese silicide alloy of nominal composition Mn$_5$Si$_3$ has attracted renewed interest among researchers due to the recent discovery of different interesting magnetic and magnetofunctional properties, which includes magnetocaloric effect, anomalous Hall effect, inverted hysteresis behavior, thermomagnetic irreversibility, and spin fluctuation~\cite{lander, menshikov, silva, gottschilch, brown, kanani, surgers, biniskos,prb-scd}. At room temperature, this Mn$_5$Si$_3$ alloy is paramagnetic (PM) in nature with hexagonal $D8_8$ type crystal structure (space group $P6_3/mcm$) having two distinct crystallographic sites for Mn (4(d) and 6(g), popularly termed as Mn1 and Mn2)~\cite{lander, menshikov}. A decrease in sample temperature ($T$) results in two magnetic transitions, namely PM to collinear antiferromagnetic (AFM2) state at 100 K and AFM2 to noncollinear antiferromagnetic (AFM1) state at 66 K~\cite{lander, menshikov}. Both of these magnetic transitions are also associated with the change in magnetic structures. Below PM to AFM2 transition temperature, the magnetic structure of the system becomes orthorhombic (space group $Ccmm$)~\cite{silva}. On the other hand, the monoclinic structure with the noncentrosymmetric $Cc2m$ space group has been observed below AFM2 to AFM1 transition point~\cite{silva}. Apart from AFM1 and AFM2 state, another intermediate noncollinear antiferromagnetic (AFM1$^{\prime}$) state has also been reported for the present alloy by S\"urgers {\it et al.}~\cite{surgers} Such intermediate AFM1$^{\prime}$ state appears during AFM1 to AFM2 field-induced transition. Detailed neutron diffraction studies confirm the presence of multiple Mn sites with different moment values and arrangements below the magnetic transition temperatures~\cite{silva}. Such dissimilar nature of Mn-sites in magnetically ordered phases plays a pivotal role in the observation of different magnetic properties. 
\par
Our recent investigation on Mn$_5$Si$_3$ alloy reveals probably the first-ever observation of inverted hysteresis loop (IHL) in a bulk antiferromagnetic compound~\cite{prb-scd}. The isothermal arrest of intermediate AFM1$^{\prime}$ state is the crucial reason for such novel observation. Besides, isothermal as well as thermomagnetic arrest lead to a significant modification in the existing phase diagram of the alloy~\cite{prb-scd}. Now, to check the robustness of the observed IHL and arrested behavior, it is pertinent to investigate the effect of Mn-site doping in Mn$_5$Si$_3$ alloy. For the pure Mn$_5$Si$_3$ alloy, the nearest neighbor Mn-Mn distance plays a crucial role in the magnetic character of the alloy, and it is lying very near to the critical values of instability~\cite{jap-Kanani,biniskos,silva,gottschilch,brown1,brown,surgers}. Mn-site doping will directly influence the Mn-Mn distance and hence the magnetic interaction in the material. 

\par
Until now, very few doping studies have been performed on the present manganese silicide system, which only reports the effect of doping on the magnetic transition temperatures and fields~\cite{songlin,jap-Kanani}. No detailed magnetic investigations (both macroscopic and microscopic) have yet been performed on such doped alloys. In the present work, we doped both smaller size Ni and larger size Cr atoms in the Mn-site of the Mn$_5$Si$_3$ alloy with the motivation of investigating the effects of such doping on the recently observed unique properties, such as IHL and thermomagnetic irreversibility. Experimental outcomes indicate a monotonous decrement of AFM1$^{\prime}$ phase fraction with increasing doping concentrations, and for 4\% doping, alloys are found to show field-induced transition directly from AFM1 to AFM2 phase (without intermediate AFM1$^{\prime}$ phase). A Gradual decrease of such intermediate AFM1$^{\prime}$ phase with doping affects the IHL and thermomagnetic irreversibility present in this system and eventually vanishes both for 4\% Ni and Cr doping cases.

\begin{figure*}[t]
\centering
\includegraphics[width = 17 cm]{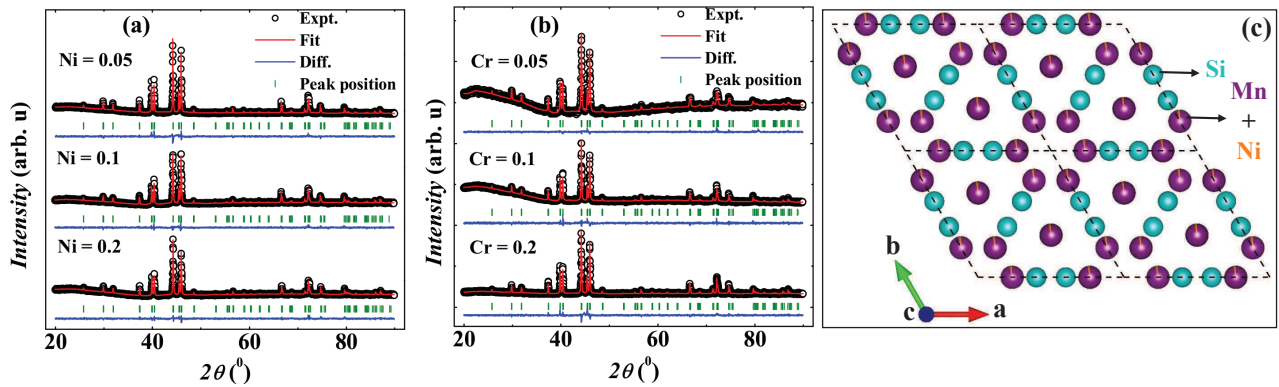}
\caption{(Color online) Room temperature x-ray powder diffraction pattern along with the Rietveld fitted curves, difference patterns, and Bragg’s peak positions of Ni and Cr-doped alloys are plotted in (a) and (b) respectively. Partial occupancies between Mn and Ni/Cr at all the parent Mn positions were considered for the best fitting of the diffraction patterns. (c) illustrates the $c$-axis projected structure of the 4\% Ni-doped alloy ($x$ = 0.2). The cyan balls represent Si atoms, whereas mixed colored balls show Mn (violet) and Ni (orange) as a combination.}
\label{xrd}
\end{figure*}

\section{Experimental details}
Ni and Cr-doped alloys of nominal compositions Mn$_{5-x}$A$_x$Si$_3$ ($x$ = 0.05, 0.1, 0.2 and A = Ni, Cr) were prepared by arc melting the constituent elements in an inert argon atmosphere using a Centorr make tri-arc furnace. Elements with purity better than 99.9\% were used for sample preparation. To ensure the homogeneity, all the alloys were remelted four times during the preparation. Finally, ingots were sealed in an evacuated quartz capsule and annealed at 900$^{\circ}$C for a week followed by rapid quenching in ice water. The room temperature x-ray powder diffraction (XRD) technique was adopted for the structural characterization of the prepared alloys. XRD patterns for the studied alloys were recorded in a Bruker AXS diffractometer (D8-Advance), where Cu-K$_{\alpha}$ radiation was used as a probe for x-ray diffraction measurements.  Commercial cryogen-free high magnetic field system from Cryogenic Ltd., UK (temperature range 2-300 K and magnetic field range 0-150 kOe), equipped with vibrating sample magnetometer (VSM), was used for dc magnetic measurements. For better temperature stability, He-exchange gas was used, and the observed fluctuation in the sample temperature was about 15 mK. During dc magnetic measurements, true zero-field cooled (ZFC) conditions were obtained by following the methods mentioned in Das {\it et al.}~\cite{prb-scd}.

\begin{figure}[t]
\centering
\includegraphics[width = 8.5 cm]{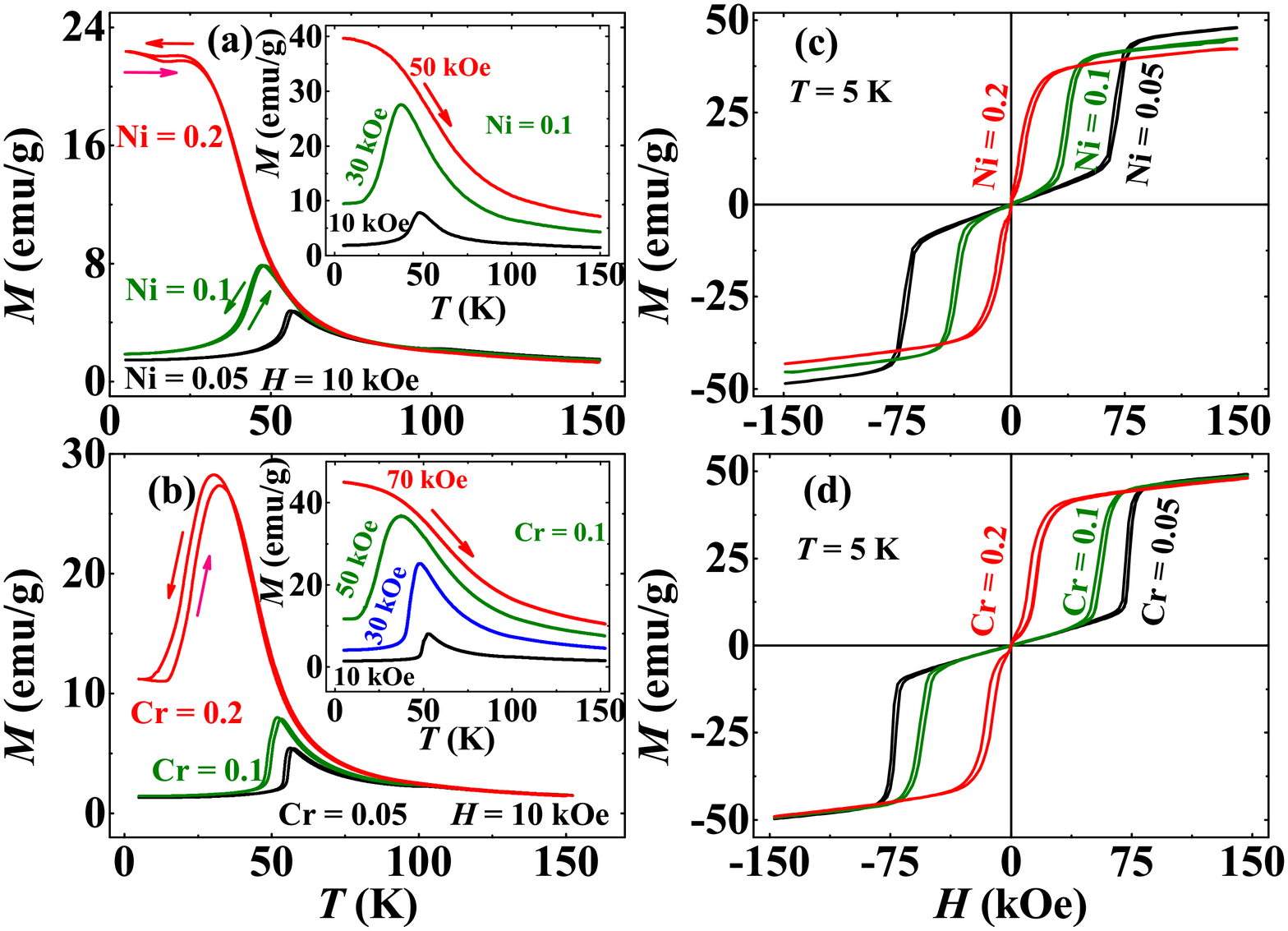}
\caption{(Color online) Main panels of (a) and (b) depict the temperature ($T$) variation of dc magnetization ($M$) in the presence of 10 kOe of an external magnetic field ($H$) in field cooling (FC) and field cooled heating (FCH) protocols for Ni and Cr-doped alloys respectively. Insets of (a) and (b) show iso-field $M(T)$ in FCH protocol at different constant $H$ for 2\% Ni and Cr-doped alloys ($x$ = 0.1) respectively. Isothermal $M(H)$ data recorded at 5 K for all Ni and Cr-doped alloys in ZFC conditions are plotted in (c) and (d), respectively.}
\label{mtmh}
\end{figure}

\begin{figure*}[t]
\centering
\includegraphics[width = 17 cm]{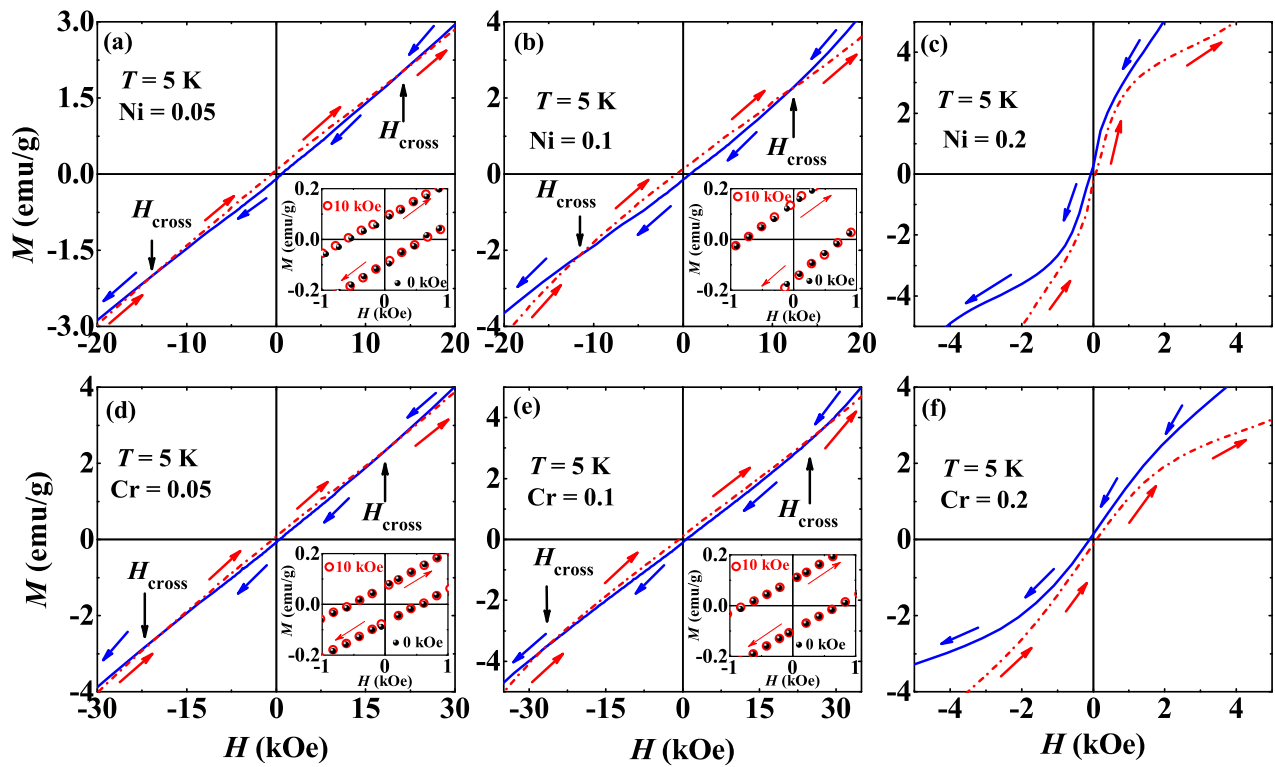}
\caption{(Color online) Restricted region of the isothermal $M(H)$ curves recorded in ZFC condition at 5 K are plotted in the main panels for all the doped alloys ((a), (b), and (c) for Ni-doped alloys and (d), (e), and (f) for Cr-doped alloys). Arrows indicate a crossing of increasing and decreasing field line curves. Insets of (a), (b),(d) and (e) depict the restricted view of zero-field cooled (black solid ball symbols) and 10 kOe field cooled (red circle symbols) isothermal $M(H)$ data at 5 K for Ni ($x$ = 0.05, and 0.1) and Cr-doped ($x$ = 0.05, and 0.1) alloys respectively.}
\label{ihl}
\end{figure*}

\section{Experimental Results}
The room temperature powder XRD patterns for the doped alloys indicate that all of them crystallize in hexagonal structure with space group $P6_3/mcm$. The diffraction patterns were analyzed by the Rietveld refinement method using the fullprof software package. XRD patterns, along with Rietveld fitted curves, difference patterns, and Bragg's peak positions are plotted in fig.~\ref{xrd}(a) and (b). For the best fitting of the diffraction patterns, we have considered partial occupancy between Mn and Ni/Cr at all the Mn positions. No impurity peak has been observed in any of the prepared alloys. Ni-doping results in a decrease in lattice volume, whereas a significant increase in lattice volume has been found for Cr-doped alloys. Variations of lattice volume with doping concentration are listed in table 1. Using Rietveld refinement parameters, we prepared the structure of 4\% Ni-doped ($x$ = 0.2) alloy and $c$-axis projected structure of the same is shown in fig.~\ref{xrd}(c).

\begin{table}
\centering
\begin{tabular}{c|c|c|c}
\hline
\hline
Sample  & Lattice  & Moment at 5 K  &   AFM2-AFM1 \\
Name & Volume & in $H$ = 150 kOe & transition  \\
& & & temperature ($T_{N1}$) \\
&\AA$^3$ & (emu/g) & (K)  \\
\hline
Mn$_{4.95}$Ni$_{0.05}$Si$_3$ & 199.09(9) & 48.03   &  56.52 \\
Mn$_{4.9}$Ni$_{0.1}$Si$_3$ & 199.01(3)  & 45.04   &  47.96 \\
Mn$_{4.8}$Ni$_{0.2}$Si$_3$ & 198.51(5)  & 42.27  &  24.16 \\
Mn$_{4.95}$Cr$_{0.05}$Si$_3$ & 199.33(1)  & 49.14  &  57.41 \\
Mn$_{4.9}$Cr$_{0.1}$Si$_3$ & 199.43(2)  & 48.61  &  53.11 \\
Mn$_{4.8}$Cr$_{0.2}$Si$_3$ & 199.58(5)  & 47.99  &  31.21 \\
\hline
\hline
\end{tabular}
\caption{Variation of lattice volume, moment at 5 K (in the presence of $H$ = 150 kOe), and AFM2 to AFM1 transition temperatures for the Ni and Cr-doped alloys.}
\end{table}

\par
Now, let us concentrate on the temperature ($T$) variation of the dc magnetization ($M$) data. The field cooling (FC) and field-cooled heating (FCH) dc $M$ data, recorded in the presence of 10 kOe of an external magnetic field ($H$), for Ni and Cr-doped alloys are plotted in fig.~\ref{mtmh} (a) and (b) respectively. These iso-field $M(T)$ data indicate a strong influence of doping on the transition temperatures and moment values. A significant decrease in the AFM1 to AFM2 transition temperatures ($T_{N1}$) has been observed in doped alloys. In the case of Ni-doping, the reduction of such transition $T$ is about 41 K for 4\% of doping ($x$ = 0.2). On the other hand, a similar percentage of Cr-doping results in a 35 K decrease in $T_{N1}$. Variation of $T_{N1}$ with doping concentration is shown in table 1. Interestingly, PM to AFM2 transition temperature ($T_{N2}$), determined by differentiating the $M(T)$ data, remains almost unchanged at 100 K with doping. The observed effect of doping is consistent with the previously reported Ni and Cr-doped Mn$_5$Si$_3$ alloys.~\cite{jap-Kanani} The presence of thermal hysteresis between FC and FCH $M(T)$ data in all Ni and Cr-doped alloys around the $T_{N1}$ confirm the first-order nature of the transition. Similar thermal hysteresis has also been observed in pristine Mn$_5$Si$_3$ alloy~\cite{prb-scd}. A significant increase in dc $M$ value for both Ni and Cr-doped alloys has been noted below $T_{N2}$ in the presence of 10 kOe of external $H$. We have also recorded iso-field $M(T)$ data in the FCH protocol at different applied $H$. Effects of increasing $H$ on $M(T)$ data of the doped alloys are found to be similar to the results observed for pure Mn$_5$Si$_3$ alloy. For both pure and doped cases, the $T_{N1}$ is shifted towards lower $T$ with increasing $H$, whereas, $T_{N2}$ remains unchanged. The observed behavior is because the external $H$ prefers the AFM2 phase over the AFM1 phase. Some representative $M(T)$ curves for both 2\% Ni and Cr-doped alloys are plotted in the insets of fig.~\ref{mtmh} (a) and (b) respectively.

\begin{figure*}[t]
\centering
\includegraphics[width = 17 cm]{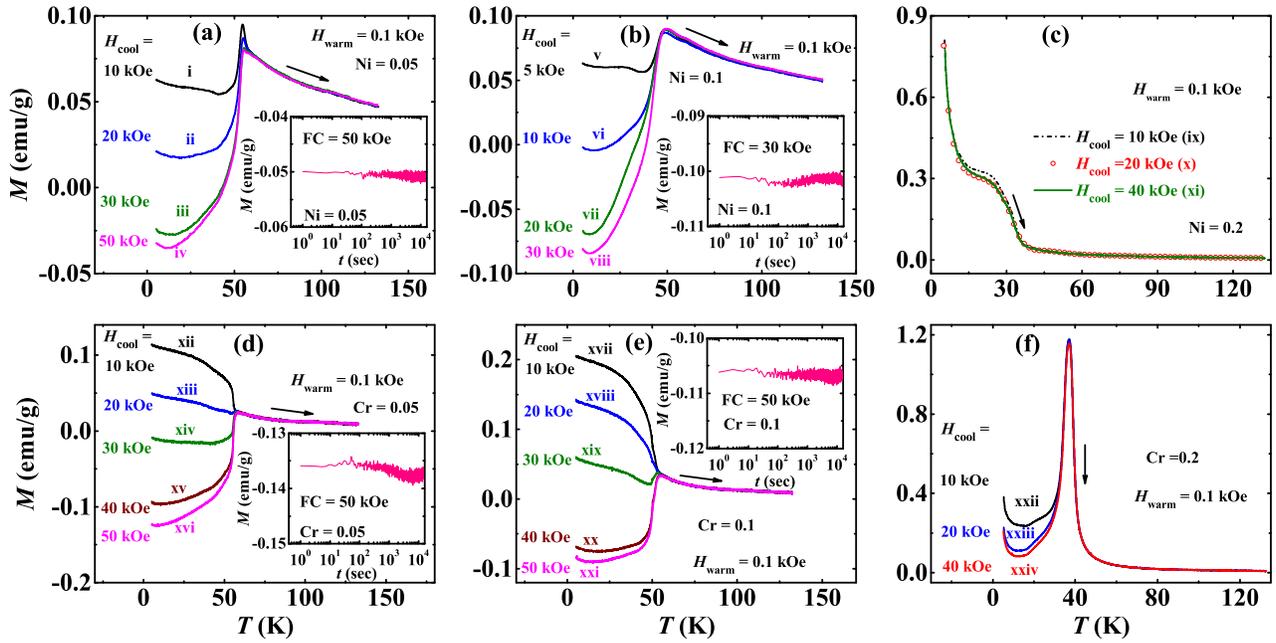}
\caption{(Color online) Main panels depict the magnetization ($M$) recorded as a function of temperature ($T$) while heating in 0.1 kOe of external magnetic field ($H$) (a) for 1\% Ni-doped alloy ($x$ = 0.05) after being cooled in $H_{\rm cool}$ = (i) 10, (ii) 20, (iii) 30 and (iv) 50 kOe; (b) for 2\% Ni-doped alloy ($x$ = 0.1) after being cooled in $H_{\rm cool}$ = (v) 5, (vi) 10, (vii) 20 and (viii) 30 kOe; (c) for 4\% Ni-doped alloy ($x$ = 0.2) after being cooled in $H_{\rm cool}$ = (ix) 10, (x) 20, and (xi) 40 kOe; (d) for 1\% Cr-doped alloy ($x$ = 0.05) after being cooled in $H_{\rm cool}$ = (xii) 10, (xiii) 20, (xiv) 30, (xv) 40 and (xvi) 50 kOe; (e) for 2\% Cr-doped alloy ($x$ = 0.1) after being cooled in $H_{\rm cool}$ = (xvii) 10, (xviii) 20, (xix) 30, (xx) 40 and (xxi) 50 kOe; (f) for 4\% Cr-doped alloy ($x$ = 0.2) after being cooled in $H_{\rm cool}$ = (xxii) 10, (xxiii) 20, and (xxiv) 40 kOe. Time ($t$) evolution of dc magnetization ($M$) recorded at 5 K in FC protocol are plotted in the insets of (a), (b), (d) and (e) for 1\% Ni-doped, 2\% Ni-doped, 1\% Cr-doped and 2\% Cr-doped alloys respectively. For FC protocol, field was applied during cooling and measurement was carried out immediately after removing the field at 5 K.}
\label{vmt}
\end{figure*} 

\par
To shed more light on the effect of Mn-site doping on the field-induced AFM1 to AFM2 transition via intermediate AFM1$^{\prime}$ phase, one of the key observation of undoped Mn$_5$Si$_3$, we recorded isothermal $M(H)$ data between $\pm$150 kOe of applied $H$ at 5 K for all the doped alloys in true ZFC condition. (see fig.~\ref{mtmh} (c) \& (d) for Ni and Cr-doped alloys respectively). Such $M(H)$ data indicate that unlike stoichiometric Mn$_5$Si$_3$ alloy, determining the signature of AFM1 to intermediate AFM1$^{\prime}$ in the form of slope change is getting harder as the critical field to reach the AFM2 phase decreases with increasing doping concentration. A faster decrease in the AFM2 transition field has been observed for Ni-doped alloys than the Cr-doped alloys. As the AFM2 critical transition field is quite different for different doping concentrations, we observed a significant difference in iso-field $M(T)$ data recorded at 10 kOe of external $H$. On the other hand, at a high-$H$ region (after reaching the AFM2 phase), a small decrease in dc $M$ value with increasing doping concentration has been noticed. It is also to be noted here that, like different transition temperatures and critical transition fields, Ni-doping also results in a faster decrease of dc $M$ value in AFM2 state at 5 K than the Cr-doping. Moment values in the presence of 150 kOe of $H$ at 5 K for all doped alloys are listed in table 1.

\begin{figure*}[t]
\centering
\includegraphics[width = 8 cm]{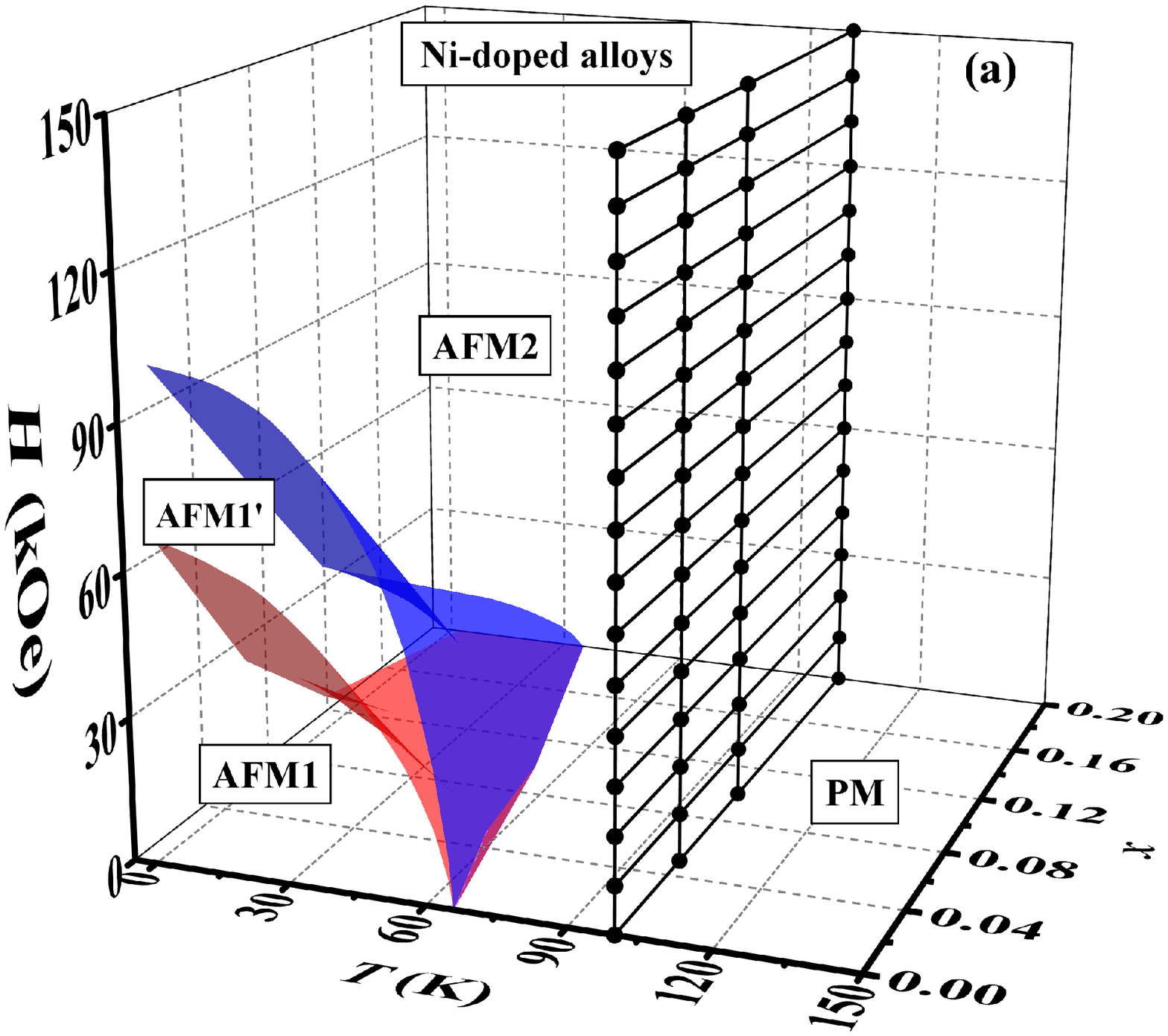}
\includegraphics[width = 8 cm]{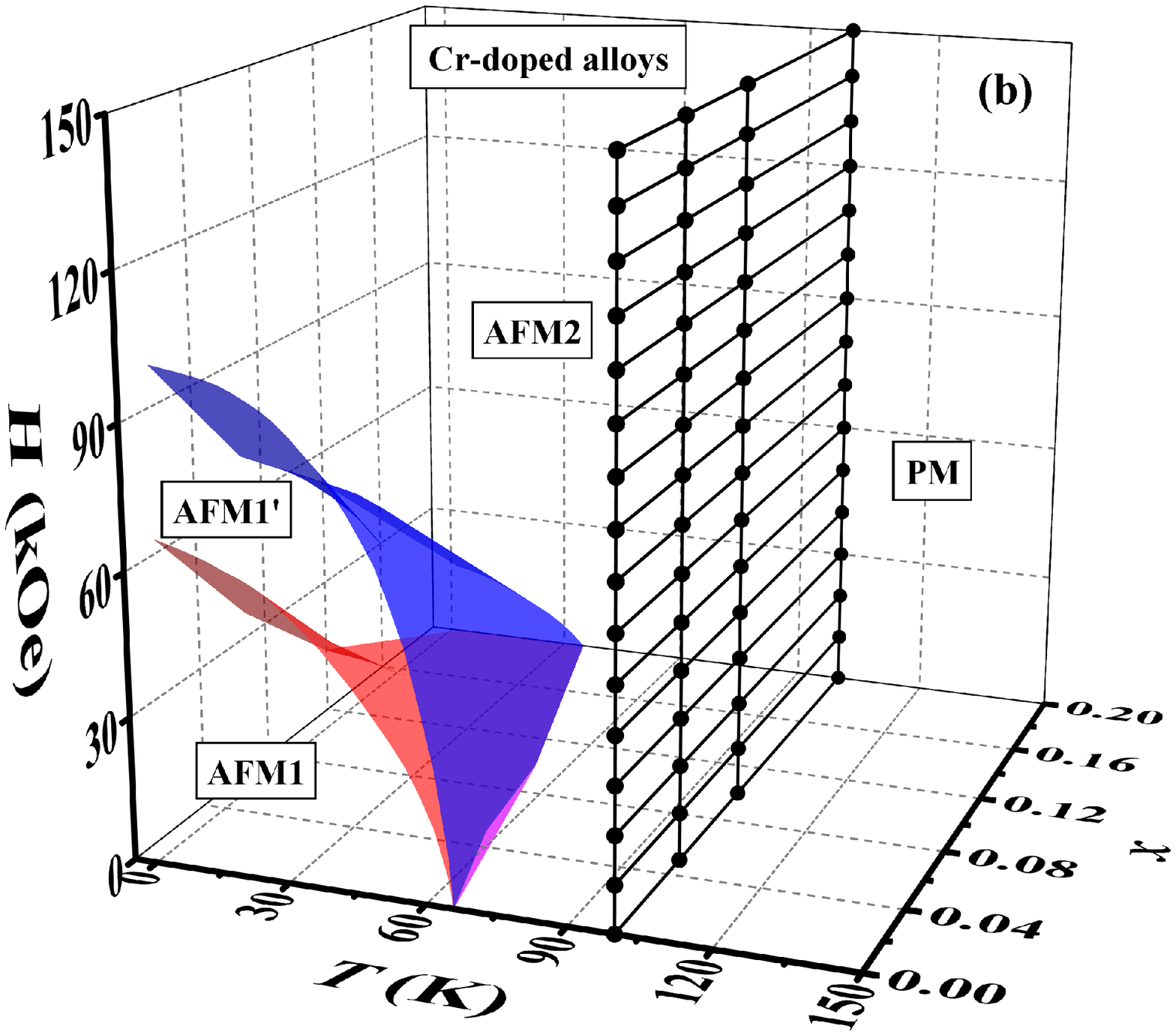}
\caption{(Color online) Field-temperature-doping concentration ($H-T-x$) phase diagrams during isothermal field application after zero-field cooling for Ni and Cr-doped alloys are depicted in (a) and (b) respectively. Red and blue colored surfaces indicate the AFM1-AFM1$^{\prime}$ and AFM1$^{\prime}$-AFM2 transition fields for different doping concentrations, respectively.}
\label{ph}
\end{figure*} 

\par
As IHL behavior, with positive coercive field and negative remanence, is one of the unique properties of pure Mn$_5$Si$_3$ alloy, it is pertinent to check the effect of doping on IHL behavior. Such investigation will also shed more light on the robustness of the IHL behavior observed in pristine Mn$_5$Si$_3$ alloy. Besides, IHL also confirms the presence of AFM1$^{\prime}$ phase in these alloys as the isothermally arrested AFM1$^{\prime}$ phase is responsible for such unusual IHL behavior~\cite{prb-scd}. Though IHL behavior is reported to be visible in exchange-coupled multilayers, hard/soft multilayers, single domain particles with competing anisotropies, and in some bulk ferrimagnet, observation of IHL in a bulk antiferromagnetic compound is infrequent (pure Mn$_5$Si$_3$ is probably the only example)~\cite{oshea, aharoni, takanashi, wn, valvidares, demirtas, kim, ohkoshi, santos, nam, geshev, song, banerjee, prb-scd, ziese}. A closure look at the $M(H)$ isotherms of the presently studied doped alloys, recorded at 5 K in actual ZFC condition, reveal that IHL is visible for both Ni and Cr-doped alloys in the low doping concentration region (for $x$ = 0.05 and 0.1; see the restricted part of the 5 K $M(H)$ isotherms for the doped alloys, plotted in the main panels of fig.~\ref{ihl} (a), (b), (d) and (e). Here we have recorded all $M(H)$ isotherms between $\pm$150 kOe). For 4\% doping (both by Ni and Cr), IHL behavior found to vanish, and conventional hysteresis loops have been observed (see fig.~\ref{ihl} (c) and (f)). The presence of exchange bias in polycrystalline alloys may sometimes give rise to IHL~\cite{ziese,jap-Zheng1}. To probe the valid reason behind the observation of IHL behavior in the presently studied alloys, we have checked the presence of any exchange bias effect in the doped alloys showing IHL by recording isothermal $M(H)$ data in 10 kOe FC condition (see insets of fig.~\ref{ihl} (a), (b), (d) and (e)).  Like the undoped alloy, we have not observed any shift in the center of the hysteresis loops for any alloy, and the FC $M(H)$ isotherms exactly matches the $M(H)$ curves recorded in ZFC condition. Such behavior confirms the role of isothermally arrested AFM1$^{\prime}$ state behind the IHL behavior.

\par
Apart from the unusual IHL properties, thermomagnetic irreversibility in $M(T)$ data is another important effect associated with the undoped Mn$_5$Si$_3$ alloy~\cite{prb-scd}. In our previous work, we have identified the thermomagnetic arrest of the AFM1$^{\prime}$ phase, which plays the key role for thermomagnetic irreversibility behaviors (we have ruled out the possibility of any arrested AFM2 state in our previous work)~\cite{prb-scd}. The presently studied doped alloys allow us to strengthen further our claim about the presence of arrested AFM1$^{\prime}$ phase. To probe the existence of such thermomagnetically arrested AFM1$^{\prime}$ phase, we recorded $T$ variation of $M$ for all the doped alloys using one of the protocols that have been used for the undoped Mn$_5$Si$_3$ alloy. Some of the representative curves for the doped alloys are depicted in the main panels of fig.~\ref{vmt} (a)-(f). Here we cooled the doped alloys from 150 K (well above the $T_{N2}$) in the presence of different applied cooling fields ($H_{\rm cool}$) and dc $M$ was recorded as a function of $T$ during heating in the presence of 100 Oe of applied field ($H_{\rm warm}$). Such protocols for $M(T)$ measurements with different cooling and heating fields are commonly used by the different groups to address the thermomagnetically arrested states in materials having first-order transitions~\cite{chatterjee, pramanick, dutta, chaddah, rawat, tokura}. For $x$=0.05 and 0.1 alloys (for both Ni and Cr-doping cases), the magnitude of dc $M$ below the $T_{N1}$ depend strongly on the $H_{\rm cool}$ (see fig.~\ref{vmt} (a), (b), (d), and (e)). An increase in the strength of $H_{\rm cool}$ results in a decrease in moment value, and eventually, it becomes negative. The critical values of $H_{\rm cool}$, for which the moments become negative, strongly depends on the doping concentration. A monotonic decrease in such critical values of $H_{\rm cool}$ has also been observed with increasing doping concentration. The magnitude of this critical field gives an idea about the AFM1-AFM1$^{\prime}$ field-induced transition, as we have identified in our previous work that the negative moment appears only if the cooling field is more than the AFM1-AFM1$^{\prime}$ transition field~\cite{prb-scd}. Observation of such irreversibility in $M(T)$ behavior and negative value of dc $M$ even in the presence of positive warming field are only possible for the presently studied alloys if there exists some thermomagnetically arrested AFM1$^{\prime}$. On the other hand, closure look to the $M(T)$ behavior with different cooling and heating fields for $x$ = 0.2 Ni-doped alloy indicates that all the curves follow the almost identical path even below the $T_{N1}$ (see fig.~\ref{vmt} (c)). For the 4\% Cr-doped alloy ($x$ = 0.2), though a very small difference in dc $M$ value at low temperatures (below $T_{N1}$) has been observed, but we failed to observe any negative value of dc $M$ in any protocol (see fig.~\ref{vmt} (f)). Such observations indicate the absence of any AFM1$^{\prime}$ state in 4\% Ni-doped alloy, whereas, a small amount of AFM1$^{\prime}$ and hence arrested AFM1$^{\prime}$ phase is present for 4\% Cr doped alloy but such a small amount of arrested AFM1$^{\prime}$ phase is not sufficient to induce IHL and negative value of $M$ below $T_{N1}$. The thermomagnetically arrested state observed for different materials is, in general, metastable in nature, and thermal energy (for $T\neq 0$) can assist those metastable systems in evolving into the equilibrium configuration. As a result of that, such materials show significantly large relaxation behavior~\cite{chatterjee}. Therefore, it is pertinent to see the time evolution of the thermomagnetically arrested state of the presently studied alloys. In the present work, we used the FC protocol for relaxation measurements. First, the sample was cooled in the presence of an external magnetic field $H_{\rm cool}$ from 150 K to 5 K. After reaching 5 K the field was removed and the dc $M$ was recorded as a function of time ($t$) (see insets of fig.~\ref{vmt} (a), (b), (d), and (e)). Depending on the magnitude of the negative moment observed during the $M(T)$ measurements, recorded under different cooling and warming field protocols, we have selected the value of $H_{\rm cool}$. The nature of the relaxation observed for both Ni and Cr doped alloys (only for $x$ = 0.05 and 0.1 concentrations) are found to be very stable (only a sluggish change in values towards the negative direction). Such observations confirm that the arrested AFM1$^{\prime}$ states are equally stable in doped alloys as it was observed for the pristine Mn$_5$Si$_3$ alloy. Relaxation measurement for 4\% doped alloys (both Ni and Cr doping cases) were not performed because of the absence of significant thermomagnetic irreversibility in $M(T)$ data.

\section{Discussions \& conclusions}
The present study of Mn-site doped Mn$_5$Si$_3$ alloys, based on structural and magnetic investigations, unveils two crucial features: (i) decrease in AFM1$^{\prime}$ phase fraction with increasing doping concentration; (ii) gradual disappearance of IHL and thermomagnetic irreversibility behavior with increasing doping concentration. Neutron diffraction studies on the undoped Mn$_5$Si$_3$ alloy confirmed that among the two Mn-sites, Mn2 splits into two independent sites (Mn21 and Mn22) along with a change in magnetic structure from hexagonal to orthorhombic below 100 K~\cite{silva}. Interestingly, out of the three Mn-sites present below 100 K, only the Mn22 site show ordered AFM moment~\cite{silva,gottschilch,brown1,brown,surgers,biniskos}. Due to the small separation between Mn1-sites ($\sim c/2$), it failed to stabilize AFM2 ordering~\cite{silva,gottschilch,brown1,brown,surgers,biniskos}. On the other hand, the triangular arrangement of Mn21 with Mn22 sites gives rise to magnetic frustration and leads to the collapse of Mn21 moment~\cite{silva,gottschilch,brown1,brown,surgers,biniskos}. Further decrease in sample temperature results in a sudden increase in lattice parameter $c$ below AFM2-AFM1 transition point (associated with), and AFM configuration stabilizes in Mn1-sites~\cite{silva,gottschilch,brown1,brown,surgers,biniskos}. It is clear that the Mn-Mn distance in this Mn$_5$Si$_3$ alloy is very near to its critical value, and a small change in Mn-Mn separation will significantly affect the magnetic configuration of the system~\cite{silva,gottschilch,brown1,brown,surgers,biniskos}. In the present work, we have altered this Mn-Mn distance by doping Ni and Cr in the Mn sites. As the Ni-atom size $<$ Mn-atom size $<$ Cr atom size, unit cell volume, and hence Mn-Mn distance decreases with Ni-doping, whereas Cr-doping results in an increase of such parameters. Interestingly for both the cases, we observe a lowering of $T_{N1}$. This indicates that a small change in Mn-Mn distance (both increase and decrease) triggers a significant impact on the magnetic ordering of the alloy. Notably, the reduction in $T_{N1}$ for Cr-doped alloys is found to be slow compare to the Ni-doped cases. A shift in $T_{N1}$ towards lower $T$ is also related to the decrease in lattice parameters of the materials in the presence of external $H$. Neutron diffraction data of pure Mn$_5$Si$_3$ alloys in the presence of an external $H$ confirms such a decrease in lattice parameters~\cite{gottschilch}. Besides, the field-induced transition from AFM1 to AFM2 phase via AFM1$^{\prime}$ phase also becomes more natural and occurs at much lower fields compare to the pristine Mn$_5$Si$_3$ alloys and the small perturbation of Mn-Mn distance in the form of Mn-site doping is playing the pivotal role. AFM1$^{\prime}$ phase has noncollinear arrangements of Mn22 atoms, but the Mn1 site moment vanishes again due to the reduction of lattice parameters (and hence Mn-Mn distance) in the presence of $H$.

\par
The gradual disappearance of IHL and thermomagnetic irreversibility, the two key observations of undoped Mn$_5$Si$_3$ alloys, are the most critical effects of Mn-site doping. In our previous work on pristine Mn$_5$Si$_3$ alloy, we have identified that the isothermally and thermomagnetically arrested of AFM1$^{\prime}$ phases are responsible for unusual IHL and thermomagnetic irreversibility behaviors respectively~\cite{prb-scd}. The AFM1$^{\prime}$ phases is an intermediate metastable phase that appears only during field-induced AFM1-AFM2 transition. The appearance of an intermediate metastable state during a first-order field-induced transition is not uncommon. During such a field-induced transition, it is often difficult for the system to overcome the substantial energy barrier and reach a more stable configuration directly. Instead, it proceeds through easily accessible local minima. Such behavior has already been observed for Heusler based shape memory alloys~\cite{chatterjee2}. In the present case, the doping of foreign elements in the Mn-site reduces the energy barrier between AFM1 and AFM2 states, along with the reduction of intermediate states (local minima). As a result, with a gradual increase of doping concentration, the system starts to reach a more stable configuration (AFM2 for the present case) in the presence of a magnetic field directly instead of going through the intermediate metastable state (AFM1$^{\prime}$ for the present case). The gradual disappearance of AFM1$^{\prime}$ phase with increasing doping concentration directly affects the unusual IHL and thermomagnetic irreversibility properties, and such properties eventually vanish for 4\% doping.

\par
Based on the experimental observations, we prepared the $H-T-x$ phase diagram for the presently studied doped alloys (see fig.~\ref{ph} (a) and (b) for Ni and Cr-doped alloys respectively). Here we have only considered the isothermal field application situation after true ZFC (out of the four different conditions explored in our previous work). All the critical fields were determined by differentiating the isothermal $M(H)$ curves recorded at different constant $T$ (not shown here). On the other hand, PM-AFM2 transition temperatures were determined from the iso-field $M(T)$ data recorded at different constant $H$. The blue colored surface indicates the AFM2 to AFM1$^{\prime}$ transition field, whereas, AFM1 to AFM1$^{\prime}$ transition fields were indicated by the red surface. A Clear decrease in the AFM1$^{\prime}$ region has been observed with increasing doping concentration.

\par
In conclusion, the present investigation on Ni and Cr-doped Mn$_5$Si$_3$ alloys reveal the doping effects on the unusual inverted hysteresis loop and thermomagnetic irreversibility properties observed for undoped Mn$_5$Si$_3$ alloy. Further, it reconfirms the role of AFM1$^{\prime}$ phase behind such unusual features. We have also tried to prepare an $H-T-x$ phase diagram for the doped alloys. Mn-site doped Mn$_5$Si$_3$ alloys (in the low doping region; up to 2\% doping), is an excellent addition to the bulk antiferromagnetic alloys, which show IHL behavior.

\section{Acknowledgments}
SCD (IF160587) and NK would like to thank DST-India and UGC-India, respectively, for their fellowship.


%

\end{document}